%% file: main.tex
\documentclass[preprint,onecolumn,aps,prd,groupedaddress,nofootinbib,showpacs]{revtex4}
\usepackage{graphicx}
\graphicspath{{fig/}} \DeclareGraphicsExtensions{.eps}
\usepackage{amsmath}
\usepackage{bm}
\usepackage{slashed}
\usepackage[hypertex]{hyperref}

\def\a{\alpha}

\def\e{\epsilon}
\def\g{\gamma}
\def\s{\sigma}

\newcommand{\jpsi}{{J/\psi}}

\newcommand{\state}[4]{{^#1\hspace{-0.6mm}#2_{#3}^{[#4]}}}

\newcommand\CScSa{\state{3}{S}{1}{1}}

\newcommand\COcPz{\state{3}{P}{0}{8}}

\newcommand\mo{{\mathcal O}}

\newcommand{\LDME}[2]{\langle\mo^{#1}(#2)\rangle}
\newcommand\mops{\LDME{\jpsi}{\CScSa}}

\begin{document}

\preprint{
          \begin{tabular}{l}
          BNL-98309-2012-JA
          \end{tabular}
          }

\title{General Form of $s$, $t$, $u$ Symmetric Polynomial and Heavy Quarkonium physics}

\author{Yan-Qing Ma}
\email{yqma@bnl.gov} \affiliation{Physics Department, Brookhaven
National Laboratory, Upton, NY 11973, USA}

\date{\today}

\begin{abstract}

Induced by three gluons symmetry, Mandelstam variables $s$, $t$, $u$
symmetric expressions are widely involved in collider physics,
especially in heavy quarkonium physics. In this work we study
general form of $s$, $t$, $u$ symmetric polynomials, and find that
they can be expressed as polynomials where the symmetry is manifest.
The general form is then used to simplify expressions which
asymptotically reduces the length of original expression to
one-sixth. Based on the general form, we reproduce the exact
differential cross section of $J/\psi$ hadron production at leading
order in $v^2$ up to four unknown constant numbers by simple
analysis. Furthermore, we prove that differential cross section at
higher order in $v^2$ is proportional to that at leading order. This
proof explains the proportion relation at next-to-leading order in
$v^2$ found in previous work and generalizes it to all order.

\end{abstract}
\pacs{12.38.Bx, 14.40.Pq}
\maketitle

\section{INTRODUCTION}\label{sec:introduction}

Quantum chromodynamics (QCD) is currently believed to be the
fundamental theory of the strong interaction. Thanks to asymptotic
freedom and
factorization~\cite{Collins:1985ue,Bodwin:1984hc,Collins:1989gx,Qiu:1990xy,Bodwin:1994jh,Kang:2011mg}
properties, application of QCD to a physical process can be
factorized into convolution of nonperturbative, but universal, long
distance matrix elements with perturbative calculable
infrared-safe~\cite{Sterman:1995fz} short distance coefficients.
Short distance coefficients at each order in perturbative expansion
can be typically understood as the differential cross section of two
partons scattering to produce $n$ partons. This $2\to n$ parton
level process is conveniently obtained from the process $0 \to 2+n$
through crossing. Although partons can be either gluon or
(anti)quark, in many cases these $2+n$ partons are mainly gluons
which are the gauge bosons of QCD. Therefore, the identity property
between gluons results in a large symmetry for these processes.

Considering the specific type of processes $0\to g(k_1) + g(k_2)+
g(k_3) + H(P)$, where $k_i$ are momentum of each gluons, $P$ is the
total momentum of a ``cluster'' $H$ which includes one or more
partons. 
The ``cluster" here means the differential cross section of the
process should be sensitive only to the total momentum $P$ of the
cluster, but insensitive to the detail within the cluster. Momentum
conservation gives
\begin{align}
k_1^{\mu}+k_2^{\mu}+k_3^{\mu}+P^{\mu}=0\,,
\end{align}
with $k_i^2=0$ and $P^2=M^2$, where $M$ is the invariant mass of
cluster $H$. The associated Mandelstam variables are given by
\begin{subequations}
\begin{align}
s&=(k_1+k_2)^2=(P+k_3)^2\,,\\
t&=(k_2-k_3)^2=(P-k_1)^2\,,\\
u&=(k_3-k_1)^2=(P-k_2)^2\,,
\end{align}
\end{subequations}
with $s+t+u=M^2$. Symmetry between the three gluons implies that the
following function is symmetric under the exchange of $s$, $t$, and
$u$:
\begin{align}
F(M^2,s,t,u)={\overline\sum} |A|^2,
\end{align}
where $A$ is the Feynman amplitude of the process and
$\overline\sum$ means summation/average over spin and color of these
three gluons. Before doing any integration, $F(M^2,s,t,u)$ is a
fraction polynomial, and both its numerator and denominator are
symmetric polynomials. As a result, to study the restriction
introduced by gluon symmetry, it is equivalent to study the property
of the symmetric polynomial.

The above type of processes are widely involved in collider physics.
The simplest example is studying two jets production in hadron
colliders, and calculation of four gluons scattering is needed,
where $H$ is also a gluon. In heavy quarkonium physics, there are a
lot of processes that belong to this type, including heavy
quarkonium decays to light hadrons~\cite{Barbieri:1976fp,
Huang:1996sw, Petrelli:1997ge, He:2008xb, Fan:2009cj, Guo:2011tz}
and heavy quarkonium production in hadron colliders (see, for
example Refs. ~\cite{Kuhn:1992qw, Cho:1995ce, Beneke:1998re,
Fan:2009zq,Xu:2012am, Butenschoen:2012px, Chao:2012iv, Gong:2012ug}
and references therein),
where $H$ is a heavy quark antiquark pair with a very small relative
momentum. Hopefully, studying the property of the symmetric
polynomial will introduce rigorous restriction for these processes.

Among others, there is a very interesting finding in heavy
quarkonium physics recently that, at leading order (LO) in $\a_s$
and in the large transverse momentum $p_T$ limit, the relativistic
correction term for $\jpsi$ hadron production is proportional to the
leading term~\cite{Fan:2009zq}. The proportion behavior is
nontrivial because there is more than one parameter even in the
large transverse momentum limit. It is likely to have a symmetry to
protect this behavior, which is one of the motivations to study the
symmetry property induced by three identical gluons.

The rest of the paper is organized as follows. We study the general
form of an $s$, $t$, $u$ symmetric polynomial in
Sec.~\ref{sec:form}. We devote Sec.~\ref{sec:massless} to the
massless case and find that the polynomial can be expressed in a
form where symmetry is manifest. By explicitly constructing, we
generalize the massless result to include also massive particles in
Sec.~\ref{sec:massive}. The general form is then used to simplify
the expression in Sec.~\ref{sec:simplify}. Asymptotically, this
simplification can reduce the length of an expression to one-sixth.
In Sec.~\ref{sec:reproduce}, we use the general form to reproduce
some known results and explain the unexpected proportion behavior of
the relativistic correction of $\jpsi$ hadron
production~\cite{Fan:2009zq}. The proportion behavior is also
generalized to all order in $v^2$. Finally, we summarize the results
in this work and give an outlook for future works in
Sec.~\ref{sec:summary}.

\section{General form of $s$, $t$, $u$ symmetric polynomial}\label{sec:form}

In this section, we study the general form of a polynomial
$F_n(M^2,s,t,u)$ which is symmetric under the exchange of $s$, $t$,
and $u$. Because all terms in $F_n(M^2,s,t,u)$ should have the same
mass dimensions, a explicit subscript $n$ is attached to denote the
mass dimensions $2n$. Define three symmetric combinations:
\begin{subequations}
\begin{align}
S_1:=&s+t+u\,,\\
S_2:=&-st-tu-us\,,\\
S_3:=&stu\,.
\end{align}
\end{subequations}
We will find that $F_n(M^2,s,t,u)$ can be expressed as a polynomial
$\hat{F}_n(S_1,S_2,S_3)$. This result is not hard to understand
because the relation $s+t+u=M^2$ has already suggested that only
three variables are independent in $F_n$. What is not clear is
whether $\hat{F}_n$ is a polynomial. In fact, $s$, $t$, $u$
symmetric variables has been used for specific processes for a long
time \cite{Gastmans:1986qv,Gastmans:1987be,Meijer:2007eb}, but a
general proof is still missing. In the following, we first study the
case where $M=0$. Then the extension to the massive case is
straightforward. Main results in this section are
Eqs.~\eqref{eq:massless} and \eqref{eq:massive}.

\subsection{Massless case: $s+t+u=0$}\label{sec:massless}

Define
\begin{align}
f_n(s,t,u):=F_n(0,s,t,u)\,,
\end{align}
which is a symmetric homogeneous polynomial. The general expression
of $f_n(s,t,u)$ is
\begin{align}
f_n(s,t,u)=\sum_{i=0}^{\frac{n+1}{2}}x_i\left[ s^{n-i}(t^i+u^i)+
t^{n-i}(u^i+s^i)+u^{n-i}(s^i+t^i)\right]+s t u~
\tilde{f}_{n-3}(s,t,u)\,,
\end{align}
where $x_i$ are independent of $s$, $t$, and $u$, and
$\tilde{f}_{n-3}(s,t,u)$ is a symmetric polynomial with power $n-3$
[obviously, $\tilde{f}_{n-3}(s,t,u)=0$ if $n < 3$]. Therefore, we
need to prove only the following function is a polynomial in $S_2$
and $S_3$:
\begin{align}\label{eq:fbar}
\bar{f}_n(s,t,u)=\sum_{i=0}^{\frac{n+1}{2}}x_i\left[
s^{n-i}(t^i+u^i)+ t^{n-i}(u^i+s^i)+u^{n-i}(s^i+t^i)\right]\,.
\end{align}
We prove it recursively.

\begin{itemize}

\item $n=0$ or $1$ is trivial.

\item $n=2$:
\begin{align}
\bar{f}_2(s,t,u)=2x_0 (s^2+t^2+u^2)+ 2x_1
(st+tu+us)=2(2x_0-x_1)S_2\,.
\end{align}

\item $n\geq3$ and $n\in 2N+1$:\\
In this case, we will show that $\bar{f}_n(s,t,u)\propto s$, and
then using symmetry we can get $\bar{f}_n(s,t,u)\propto stu$, that
is, $\bar{f}_n(s,t,u)=S_3 f_{n-3}(s,t,u)$.
 Setting $s=0$, one has $t=-u$, thus,
\begin{align}
\bar{f}_n(0,-u,u)=\sum_{i=0}^{\frac{n+1}{2}}x_i\left[t^{n-i}u^i+u^{n-i}t^i\right]
=\sum_{i=0}^{\frac{n+1}{2}}x_i\left[(-1)^{n-i}+(-1)^i\right]u^n=0\,.
\end{align}
Recall the general expression of $\bar{f}_n(s,t,u)$ in
Eq.~\eqref{eq:fbar}, the above result means $\bar{f}_n(s,t,u)\propto
s$.

\item $n\geq3$ and $n \in 2N$, but $n \not\in 6N$:\\
In this case, we will show that $\bar{f}_n(s,t,u)\propto S_2$.
Solutions of
\begin{align}
\begin{cases}
S_1=0\,,\\
S_2=0\,,
\end{cases}
\end{align}
are
\begin{align}\label{eq:sol}
\begin{cases}
s=e^{\pm i \frac{2\pi}{3}}u\,,\\
t=e^{\mp i \frac{2\pi}{3}}u\,.
\end{cases}
\end{align}
Thus, equivalently, we need to show that $\bar{f}_n(s,t,u)$ vanishes
for solutions in Eq.~\eqref{eq:sol}:
\begin{align}
\begin{split}
&\bar{f}_n(e^{\pm i \frac{2\pi}{3}}u,e^{\mp i \frac{2\pi}{3}}u,u)\\
=&u^n\sum_{j=0}^{\frac{n}{2}}x_j\left[e^{\pm i
\frac{2(n-j)\pi}{3}}(e^{\mp i \frac{2j\pi}{3}}+1)+e^{\mp i
\frac{2(n-j)\pi}{3}}(e^{\pm i \frac{2j\pi}{3}}+1)+(e^{\pm
i\frac{2j\pi}{3}}+e^{\mp i \frac{2j\pi}{3}})\right]\\
=&u^n\sum_{j=0}^{\frac{n}{2}}x_j\left[e^{\pm i
\frac{2(n-2j)\pi}{3}}+e^{\pm i \frac{2(n-j)\pi}{3}}+e^{\pm i \frac{2j\pi}{3}}+c.c.\right]\\
=&2u^n\sum_{j=0}^{\frac{n}{2}}x_j\left[\cos{\frac{2(n-2j)\pi}{3}}+\cos{
\frac{2(n-j)\pi}{3}}+\cos{\frac{2j\pi}{3}}\right]\,.
\end{split}
\end{align}
Because $\cos{\frac{2j\pi}{3}}=1$ if $j\in 3N$ and
$\cos{\frac{2j\pi}{3}}=-\frac{1}{2}$ if $j\not\in 3N$,
$\cos{\frac{2(n-2j)\pi}{3}}+\cos{
\frac{2(n-j)\pi}{3}}+\cos{\frac{2j\pi}{3}}=0$ for all $j$, that is
$\bar{f}_n(e^{\pm i \frac{2\pi}{3}}u,e^{\mp i
\frac{2\pi}{3}}u,u)=0$.

\item $n\geq3$ and $n\in 6N$:\\
Define
\begin{align}
G_i(s,t,u):=s^{n-i}(t^i+u^i)+ t^{n-i}(u^i+s^i)+u^{n-i}(s^i+t^i)\,.
\end{align}
It is easy to find that
\begin{align}
\begin{cases}
G_j(0,-u,u)=2(-1)^j u^n\,,\\
G_j(e^{\pm i \frac{2\pi}{3}}u,e^{\mp i \frac{2\pi}{3}}u,u)=6e^{i
\frac{2j\pi}{3}}u^n\,.
\end{cases}
\end{align}
Thus,
\begin{align}
\begin{cases}
(-1)^{j_1}G_{j_1}(0,-u,u)-(-1)^{j_2}G_{j_2}(0,-u,u)=0\,,\\
e^{-i \frac{2{j_1}\pi}{3}}G_{j_1}(e^{\pm i \frac{2\pi}{3}}u,e^{\mp i
\frac{2\pi}{3}}u,u)-e^{-i \frac{2{j_2}\pi}{3}}G_{j_2}(e^{\pm i
\frac{2\pi}{3}}u,e^{\mp i \frac{2\pi}{3}}u,u)=0\,,
\end{cases}
\end{align}
which means
\begin{align}\label{eq:6N}
\begin{cases}
(-1)^{j_1}G_{j_1}(s,t,u)-(-1)^{j_2}G_{j_2}(s,t,u)\propto S_3\,,\\
e^{-i \frac{2{j_1}\pi}{3}}G_{j_1}(s,t,u)-e^{-i
\frac{2{j_2}\pi}{3}}G_{j_2}(s,t,u)\propto S_2\,.
\end{cases}
\end{align}
For any $j_1$, there exists a $j_2$ which guarantees the coefficient
matrix of Eq.~\eqref{eq:6N} to be nonzero. Therefore, the solution
of Eq.~\eqref{eq:6N} gives
\begin{align}
G_j(s,t,u)=S_2A_j(s,t,u)+S_3B_j(s,t,u)\,,
\end{align}
where $A_j(s,t,u)$ and $B_j(s,t,u)$ are symmetric polynomial.
Specifically, taking advantage of results in previous cases, we find
$A_j(s,t,u)\propto S_2^2$ and $B_j(s,t,u)\propto S_3$. Using this
argument recursively, one gets that $G_j(s,t,u)$ is a polynomial in
$S_2^3$ and $S_3^2$. As a result, $\bar f_n(s,t,u)$ is a polynomial
in $S_2^3$ and $S_3^2$.
\end{itemize}

Combine all possible cases above, we indeed proved that $f_n(s,t,u)$
is a polynomial in $S_2$ and $S_3$ for any $n$. More precisely, we
find
\begin{align}\label{eq:massless}
f_n(s,t,u)=\hat{f}_n(S_2,S_3)=\begin{cases}
X_{\frac{n}{6}}(S_2^3,S_3^2), n\in6N\,,\\
S_2^2S_3X_{\frac{n-7}{6}}(S_2^3,S_3^2), n\in6N+1\,,\\
S_2X_{\frac{n-2}{6}}(S_2^3,S_3^2), n\in6N+2\,,\\
S_3X_{\frac{n-3}{6}}(S_2^3,S_3^2), n\in6N+3\,,\\
S_2^2X_{\frac{n-4}{6}}(S_2^3,S_3^2), n\in6N+4\,,\\
S_2S_3X_{\frac{n-5}{6}}(S_2^3,S_3^2), n\in6N+5\,,\\
\end{cases}
\end{align}
where $X_{i}(x,y)$ is an arbitrary homogeneous polynomial in $x$ and
$y$ with power $i$. Specially, $X_{i}(x,y)=0$ if $i<0$.

\subsection{Massive case: $s+t+u=M^2$}\label{sec:massive}

When $S_1=M^2\neq0$, we will show that $F_n(M^2,s,t,u)$ can be
expressed as:
\begin{align} \label{eq:massive}
F_n(M^2,s,t,u)=\hat{F}_n(S_1,S_2,S_3)=\sum_{i=0}^n S_1^i
\hat{f}_{n-i}(S_2,S_3)\,,
\end{align}
where general form of $\hat{f}_{n-i}(S_2,S_3)$ is explicit in
Eq.~\eqref{eq:massless}. We prove Eq.~\eqref{eq:massive} by
construction.

First, the $i=0$ term in Eq.~\eqref{eq:massive} can be obtained by
setting $S_1=M^2=0$ in $F_n(M^2,s,t,u)$ using the method discussed
in Sec.~\ref{sec:massless}. Then we find
$F_n(M^2,s,t,u)-\hat{f}_{n}(S_2,S_3)$ is zero if one sets $S_1=0$,
that is, $F_n(M^2,s,t,u)-\hat{f}_{n}(S_2,S_3) \propto S_1$.
Considering that $F_n(M^2,s,t,u)$, $\hat{f}_{n}(S_2,S_3)$, and $S_1$
are symmetric polynomials,
\begin{align}
F_{n-1}(M^2,s,t,u)=\frac{F_n(M^2,s,t,u)-\hat{f}_{n}(S_2,S_3)}{S_1}
\end{align}
is also a symmetric polynomial. Applying this method repeatedly, the
result of Eq.~\eqref{eq:massive} can be achieved.

\section{Simplify expressions}\label{sec:simplify}

General forms in the last section can be used to simplify
expressions that are symmetric polynomials in $s$, $t$, and $u$ by
expressing them in terms of manifest symmetric form. In the large
$n$ limit, one finds from Eq.~\eqref{eq:massless} that the symmetric
form is a polynomial with power $\frac{n}{6}$, therefore,
asymptotically this method can reduce the length of the original
expression to one-sixth. Also, the method can be easily realized in
terms of a computer program. 
Here, we give two examples.

The first example is an ideal expression
\begin{align}
f_{42}:=\sum_{i=1}^{42}
s^i(t^{42-i}+u^{42-i})+t^i(u^{42-i}+s^{42-i})+u^i(s^{42-i}+t^{42-i})\,,
\end{align}
with $s+t+u=0$. Its explicit expression in terms of $t$ and $u$ is
\begin{align}\label{eq:eg1}
\begin{split}
f_{42}&= 2 (t^2 + t u + u^2)^3  \bigl[t^{36}+18 t^{35} u+342 t^{34}
u^2+4029 t^{33} u^3+34542
   t^{32} u^4+229536 t^{31} u^5\\
&  +1229611 t^{30}
   u^6+5455041 t^{29} u^7+20431896 t^{28} u^8+65541187
   t^{27} u^9+182032158 t^{26} u^{10}\\
&  +441440337 t^{25}
   u^{11}+940899497 t^{24} u^{12}+1771715799 t^{23}
   u^{13}+2959077438 t^{22} u^{14}\\
&   +4396930001 t^{21}
   u^{15}+5825638020 t^{20} u^{16}+6892901679 t^{19}
   u^{17}+7289748245 t^{18} u^{18}\\
&  +6892901679 t^{17}
   u^{19}+5825638020 t^{16} u^{20}+4396930001 t^{15}
   u^{21}+2959077438 t^{14} u^{22}\\
&  +1771715799 t^{13}
   u^{23}+940899497 t^{12} u^{24}+441440337 t^{11}
   u^{25}+182032158 t^{10} u^{26}\\
&  +65541187 t^9
   u^{27}+20431896 t^8 u^{28}+5455041 t^7 u^{29}+1229611
   t^6 u^{30}+229536 t^5 u^{31}\\
&   +34542 t^4 u^{32}+4029 t^3
   u^{33}+342 t^2 u^{34}+18 t u^{35}+u^{36}\bigr]\,.
\end{split}
\end{align}
As $f_{42}$ is symmetric under exchanging of $s$, $t$, and $u$, we
can express it in terms of $S_2$ and $S_3$. Specifically, we find
the result for terms within brackets of Eq.~\eqref{eq:eg1} is
\begin{align}\label{eq:eg1f}
\Bigl[\cdots\Bigr]=S_2^{18}+171 S_2^{15} S_3^2+3060
   S_2^{12} S_3^4+12376 S_2^9
   S_3^6+12870 S_2^6 S_3^8+3003
   S_2^3 S_3^{10}+91 S_3^{12}\,,
\end{align}
which is much simpler than the original expression. The result in
Eq.~\eqref{eq:eg1f} has two meanings. First, it tests the proof in
Sec.~\ref{sec:massless} to be true by explicit calculation. Second,
it shows the asymptotic behavior that our method can reduce the
length of the original expression to one-sixth.

The second example is a massive one. We take the Eq. (A5d) of
Ref.~\cite{Cho:1995ce}:
\begin{align}\label{eq:eg2}
\begin{split}
& \mathop{{\overline{\sum}}} | {\cal A}(gg \to c\bar{c}[\COcPz] g)
|^2= {5 (4\pi \a_s)^3 \over 12 M^3 \bigl[s z^2 (s-M^2)^4 (s M^2 +
z^2)^4
 \bigr]} \Bigl\{\\
&\qquad\qquad\qquad +s^2 z^4 (s^2-z^2)^4 + M^2 s z^2 (s^2-z^2)^2
(3s^2-2z^2) (2s^4 - 6 s^2 z^2 + 3 z^4) \\ & \qquad\qquad\qquad + M^4
\bigl[ 9s^{12} - 84 s^{10} z^2
  + 265 s^8 z^4
  - 382 s^6 z^6 + 276 s^4 z^8 - 88 s^2 z^{10}
  + 9 z^{12} \bigr]
\\ & \qquad\qquad\qquad - M^6 s \bigl[ 54 s^{10} - 357 s^8 z^2
 + 844 s^6 z^4 - 898 s^4 z^6 + 439 s^2 z^8
 - 81 z^{10} \bigr] \\
& \qquad\qquad\qquad + M^8 \bigl[ 153 s^{10} - 798 s^8 z^2
  + 1415 s^6 z^4
  - 1041 s^4 z^6 + 301 s^2 z^8 - 18 z^{10} \bigr] \\
& \qquad\qquad\qquad -M^{10} s \bigl[ 270 s^8 - 1089 s^6 z^2
  + 1365 s^4 z^4 - 616 s^2 z^6 + 87 z^8 \bigr] \\
& \qquad\qquad\qquad + M^{12} \bigl[ 324 s^8 - 951 s^6 z^2
  + 769 s^4 z^4
  - 189 s^2 z^6 + 9 z^8 \bigr] \\
& \qquad\qquad\qquad - 9 M^{14} s (6 s^2 - z^2) (5 s^4
  - 9 s^2 z^2 + 3 z^4)  3 M^{16} s^2 (51 s^4-59 s^2 z^2
  + 12 z^4)\\
& \qquad\qquad\qquad   - 27 M^{18} s^3 (2s^2 - z^2) + 9 M^{20} s^4
\Bigr\}\,,
\end{split}
\end{align}
which is the squared amplitude of $\COcPz$ channel for $\jpsi$
hadron production via gluon-gluon fusion. Even though a new variable
$z:=\sqrt{tu}$ is introduced, the expression is still very long.
Using our method, it can be expressed in a much more compact form.
Terms within braces of Eq.~\eqref{eq:eg2} can be expressed as
\begin{align}
\begin{split}
&\Bigl\{\cdots\Bigr\} =S_2^4 S_3^2+S_1
   S_2^2 S_3 \left(6
   S_2^3-S_3^2\right)+S_1^2 \left(9
   S_2^6+4 S_2^3 S_3^2+5
   S_3^4\right)+S_1^3 S_2 S_3
   \left(27 S_2^3-11 S_3^2\right)\\
&  2
   S_1^4 S_2^2 \left(9 S_2^3-16
   S_3^2\right)- S_1^5 S_3 \left(3
   S_2^3+13 S_3^2\right)+9 S_1^6
   S_2 \left(S_2^3-2
   S_3^2\right)-9 S_1^7 S_2^2
   S_3\,.
\end{split}
\end{align}
More compactly, ordering it according to the power of $S_3$, we have
\begin{align}
\begin{split}
&\Bigl\{\cdots\Bigr\} =9 S_1^2 S_2^4 \left(S_1^2+ S_2\right)^2 -3
S_1 S_2^2 S_3 \left(3S_1^6+ S_1^4 S_2-9 S_1^2
S_2^2-2 S_2^3\right)\\
&  - S_2 S_3^2 \left(18 S_1^6+32 S_1^4 S_2-4 S_1^2 S_2^2-
S_2^3\right) -S_1 S_3^3 \left(13 S_1^4+11 S_1^2 S_2+S_2^2\right) +5
S_1^2 S_3^4\,.
\end{split}
\end{align}
Note, although, that the simplification in this example is not as
significant as the previous one, which is because the power here is
smaller.

\section{$\jpsi$ hadron production}\label{sec:reproduce}

\subsection{$\jpsi$ hadron production at leading order in
$v^2$}\label{sec:jpsiLO}

\begin{figure}[htb!]
 \begin{center}
 \includegraphics[width=0.30\textwidth]{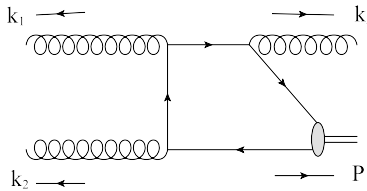}
  \caption[$\jpsi$]{One out of six Feynman diagrams for $g+g \to c\bar{c}[\CScSa] +
  g$, where all momentum are outgoing. The other five diagrams can be obtained by permutating the three
  gluons.
   \label{fig:jpsi}}
 \end{center}
\end{figure}

A well-known result in heavy quarkonium physics is the $\jpsi$
hadron production in color-singlet
model\cite{Ellis:1976fj,Carlson:1976cd,Chang:1979nn,Baier:1981uk},
one out of six Feynman diagrams for which at leading order in $\a_s$
is shown in Fig.~\ref{fig:jpsi}. Based on results in
Sec.~\ref{sec:form}, we will reproduce the behavior of the exact
result by simple analysis.

It is easy to find that denominator of the amplitude in
Fig.~\ref{fig:jpsi} is proportional to $(s-M^2)(u-M^2)$, thus, the
summation of all six diagrams is proportional to
$\left[(s-M^2)(t-M^2)(u-M^2)\right]^{-1}$. In the Feynman gauge,
squaring the summed amplitude and summing/average over polarization,
we get
\begin{align}\label{eq:Fss}
F(\CScSa)\propto\frac{F_6(M^2,s,t,u)}{\left[(s-M^2)(t-M^2)(u-M^2)\right]^2}\,.
\end{align}
Note that, an additional dimensionless factor $\mops/M^3$ and a flux
factor $1/2s$ are necessary to get the cross section of $\jpsi$
production, but for our purpose it does not matter and we will
neglect them. Before further discussion, we will prove that the
behavior of this process to the differential cross section is
$\frac{d\s}{dp_T^2}\propto M^4/p_T^8$ in large $p_T$ limit, namely
$s,~t,~u\gg M^2$. Note that, to demonstrate it, we will choose a
gauge other than Feynman gauge.

We begin with the amplitude in Fig.~\ref{fig:jpsi} (we label it with
a "$1$" to denote that it is the first diagram, while the other five
diagrams will be labeled with $2$, ..., $6$.)
\begin{align}\label{eq:A1}
\begin{split}
\mathcal{A}_1&=\text{Tr}\left[
(-\frac{\slashed{P}}{2}+m)\g^\a(\frac{\slashed{P}}{2}+m)(-i g
T^{a_3}\g^{\mu_3})\frac{i(\slashed{k}_3 +
\frac{\slashed{P}}{2}+m)}{(k_3+\frac{P}{2})^2-m^2}(-i g
T^{a_1}\g^{\mu_1})\right.\\
&\qquad\qquad\qquad\qquad\qquad\qquad\left.\frac{i(-\slashed{k}_2
-\frac{\slashed{P}}{2}+m)}{(-k_2-\frac{P}{2})^2-m^2} (-i g
T^{a_2}\g^{\mu_2}) \right]\\
&=-i2g^3m\text{Tr}\left[T^{a_1}T^{a_2}T^{a_3}\right]
\text{Tr}\left[\g^\a(\frac{\slashed{P}}{2}+m) \g^{\mu_3}
\frac{\slashed{k}_3 + \frac{\slashed{P}}{2}+m}{k_3\cdot P}
\g^{\mu_1} \frac{-\slashed{k}_2 - \frac{\slashed{P}}{2}+m}{k_2\cdot
P}\g^{\mu_2}  \right]\,,
\end{split}
\end{align}
where $\mu_i$ ($a_i$) is the spin index (color index) of the gluon
with momentum $k_i$, $\a$ is the spin index of the pair
$c\bar{c}[\CScSa]$, and $m=M/2$ is the mass of the charm quark. To
get the above result, $P^2=4m^2$ and $P^\a=0$~\footnote{This is
possible because we will contract $\mathcal{A}_1$ with the summation
of polariztion $-g^{\a\a^{\prime}}+\frac{P^\a P^{\a^\prime}}{M^2}$.}
have been used. Notice that, if one sets $m=0$, the trace of the
gamma chain in the last line in Eq.~\eqref{eq:A1} vanishes because
there are an odd number of gamma matrixes then, which implies
$\mathcal{A}_1\propto M^2$. Therefore, squaring the summation of all
six amplitudes and contracting with summations of polarizations, we
will get $\frac{d\s}{dp_T^2}\propto M^4/p_T^8$ if all denominators
do not vanish when $M\to 0$. The only possible vanishing denominator
comes from the summation of spin polarization of $c\bar{c}[\CScSa]$:
$-g^{\a\a^{\prime}}+\frac{P^\a P^{\a^\prime}}{M^2}$, where the
second term violates the above argument. Thus, we should prove that
contracting with $\frac{P^\a P^{\a^\prime}}{M^2}$ will at most give
contributions at $O(M^4)$.

Contracting the trace of the gamma chain in the last line in
Eq.~\eqref{eq:A1} with $\frac{P^\a}{M}$, we get
\begin{align}\label{eq:A1bar}
\begin{split}
\bar{\mathcal{A}}_1&=\frac{1}{M}\text{Tr}\left[\slashed{P}(\frac{\slashed{P}}{2}+m)
\g^{\mu_3} \frac{\slashed{k}_3 + \frac{\slashed{P}}{2}+m}{k_3\cdot
P} \g^{\mu_1} \frac{-\slashed{k}_2 -
\frac{\slashed{P}}{2}+m}{k_2\cdot P}\g^{\mu_2}  \right]\\
&=2~\text{Tr}\left[\slashed{P}\g^{\mu_3} \frac{\slashed{k}_3 +
\frac{\slashed{P}}{2}}{2k_3\cdot P} \g^{\mu_1} \frac{-\slashed{k}_2
- \frac{\slashed{P}}{2}}{2k_2\cdot P}\g^{\mu_2}  \right]+O(M^2)\,.
\end{split}
\end{align}
It is convenient to choose an axial gauge so that $P\cdot\e(k_i)=0~
(i=1,2,3)$, where $\e(k_i)$ is the polarization vector of gluon with
momentum $k_i$. This choice of gauge does not change our argument in
the last paragraph, because it does not introduce small
denominators. In this gauge, terms more than two $\slashed{P}$ in
Eq.~\eqref{eq:A1bar} contribute to higher order in $M^2$, thus,
\begin{align}\label{eq:A1bar2}
\begin{split}
\bar{\mathcal{A}}_1&=-2~\text{Tr}\left[\slashed{P}\g^{\mu_3}
\frac{\slashed{k}_3 }{2k_3\cdot P} \g^{\mu_1} \frac{\slashed{k}_2
}{2k_2\cdot P}\g^{\mu_2} \right]+O(M^2)\\
&=\text{Tr}\left[\g^{\mu_3} \g^{\mu_1} \frac{\slashed{k}_2
}{2k_2\cdot P}\g^{\mu_2} \right]+\text{Tr}\left[\g^{\mu_3}
\frac{\slashed{k}_3 }{2k_3\cdot P} \g^{\mu_1}\g^{\mu_2}
\right]+O(M^2)\\
&=\text{Tr}\left[\g^{\mu_3} \g^{\mu_1} \frac{\slashed{k}_2
}{u}\g^{\mu_2} \right]+\text{Tr}\left[\g^{\mu_3} \frac{\slashed{k}_3
}{s} \g^{\mu_1}\g^{\mu_2} \right]+O(M^2)\,.
\end{split}
\end{align}
By circulating the three gluons, we get two other amplitudes which
are proportional to $\text{Tr}\left[T^{a_1}T^{a_2}T^{a_3}\right]$,
with
\begin{subequations}\label{eq:Aibar}
\begin{align}
\bar{\mathcal{A}}_2&=\text{Tr}\left[\g^{\mu_1} \g^{\mu_2}
\frac{\slashed{k}_3 }{s}\g^{\mu_3} \right]+\text{Tr}\left[\g^{\mu_1}
\frac{\slashed{k}_1 }{t} \g^{\mu_2}\g^{\mu_3} \right]+O(M^2)\,,\\
\bar{\mathcal{A}}_3&=\text{Tr}\left[\g^{\mu_2} \g^{\mu_3}
\frac{\slashed{k}_1 }{t}\g^{\mu_1} \right]+\text{Tr}\left[\g^{\mu_2}
\frac{\slashed{k}_2 }{u} \g^{\mu_3}\g^{\mu_1} \right]+O(M^2)\,.
\end{align}
\end{subequations}
Hence, using the relation
$\left\{\slashed{\e}(k_i),\slashed{k}_i\right\}=0$, we have
$\bar{\mathcal{A}}_1+\bar{\mathcal{A}}_2+\bar{\mathcal{A}}_3=O(M^2)$.
Considering that the summation of the other three amplitudes has the
same behavior, we find contracting with $\frac{P^\a
P^{\a^\prime}}{M^2}$ will give contributions at $O(M^6)$. Finally,
we complete the proof that $\frac{d\s}{dp_T^2}\propto M^4/p_T^8$.

As $\frac{d\s}{dp_T^2}\propto M^4/p_T^8$ at large $p_T$ limit,
$F(\CScSa)$ and $F_6(M^2,s,t,u)$ in Eq.~\eqref{eq:Fss} must be
proportional to $M^4$. Using the results in Sec.~\ref{sec:form}, we
get the general form
\begin{align}
F_6(M^2,s,t,u)= M^4 F_4(M^2,s,t,u)=S_1^2(x_0 S_2^2 + x_1 S_1 S_3 +
x_2 S_1^2 S_2 + x_4 S_1^4)\,,
\end{align}
where $x_i~(i=0,1,2,4)$ are dimensionless numbers. The above simple
analysis indeed reproduced the behavior of the exact result (e.g.,
one can find it in~\cite{Fan:2009zq}), which expressed in our form
is
\begin{align}\label{eq:Fssfinal}
F(\CScSa)\propto\frac{S_1^2\left(S_2^2 - S_1
S_3\right)}{\left[(s-M^2)(t-M^2)(u-M^2)\right]^2}\,.
\end{align}

\subsection{Relativistic corrections for $\jpsi$ hadron production}\label{sec:jpsiNLO}

A more interesting application of the results in Sec.~\ref{sec:form}
is to understand the relativistic correction behavior of $\jpsi$
hadron production in large $p_T$ limit. We will study in the limit
$s,~t,~u\gg M^2$, where the result at leading order in $v^2$ as
shown in Sec.~\ref{sec:jpsiLO} is
\begin{align}
F(\CScSa)\propto\frac{S_1^2S_2^2}{S_3^2}\,.
\end{align}
We will demonstrate that, in the large $p_T$ limit, relativistic
correction does not change this behavior, therefore, the
relativistic correction term is proportional to the leading term.
The proof includes two steps. First, we show that the denominator of
summed amplitude is not changed by relativistic correction. Second,
we show that the large $p_T$ behavior $\frac{d\s}{dp_T^2}\propto
M^4/p_T^8$ is also not changed by relativistic correction. Combining
these two points and using the general form in Sec.~\ref{sec:form},
it is straightforward to find that the relativistic correction term
is proportional to $\frac{S_1^2S_2^2}{S_3^2}$.

Before two steps arguments, we briefly introduce the relativistic
correction. In Sec.~\ref{sec:jpsiLO}, we use the color-singlet model
to calculate the $\jpsi$ hadron production, where a charm (anti)
quark pair with a definite quantum number ($c\bar{c}[\CScSa]$) is
produced in hard collision. Furthermore, we did the nonrelativistic
approximation that the relative momentum between the $c\bar{c}$ pair
is zero there. The relativistic corrections in this paper refer to
corrections by expanding the relative momentum to higher power. For
definiteness, we denote the momentum of $c$ and $\bar c$ as
\begin{align}
\begin{cases}
p_c=\frac{P}{2}+q\,,\\
p_{\bar c}=\frac{P}{2}-q\,,\\
\end{cases}
\end{align}
where $P$ is the total momentum of $c\bar{c}$ pair and $q$ is half
of the relative momentum. As the $\jpsi$ is an $S$-wave quarkonium,
only terms with even power of relative momentum in amplitude level
contribute. In addition, the projecting amplitude to $S$ wave is
equivalent to contract coefficients of relative momentum with terms
like $-g^{\mu\nu}+\frac{P^\mu P^\nu}{M^2}$.

Now, we are ready for the first step. The only possible source that
may change the denominator of the summed amplitude is the expansion
of relative momentum for the denominator with finite relative
momentum. As an example, we study the expansion of the upper
denominator in Fig.~\ref{fig:jpsi}:
\begin{align}
\frac{1}{(k_3+\frac{P}{2}+q)^2-m^2}=\frac{1}{k_3\cdot
P}-\frac{2k_3^\mu}{(k_3\cdot P)^2}q_\mu+\cdots\,.
\end{align}
When contracting $k_3^\mu$ with $-g^{\mu\nu}+\frac{P^\mu
P^\nu}{M^2}$ to get the $S$ wave, $\frac{P^\mu P^\nu}{M^2}$ gives
the leading contribution, while $-g^{\mu\nu}$ is suppressed as $M\to
0$. Notice that contracting $k_3^\mu$ with $\frac{P^\mu P^\nu}{M^2}$
cancels the denominator $(k_3\cdot P)^{-2}$ by one power exactly,
therefore, we find that expanding this denominator to higher order
in $q$ does not change the denominator at large $p_T$ limit. This
argument can be easily extended to any denominator that expands to
any power of $q$. Thus, the first step is achieved.

Let us then argue that the large $p_T$ behavior
$\frac{d\s}{dp_T^2}\propto M^4/p_T^8$ holds to all orders of
relativistic corrections, that is, terms of order $M^2/p_T^6$ vanish
at the cross section level. For a finite relative momentum, the
amplitude in Fig.~\ref{fig:jpsi} is
\begin{align}\label{eq:A1v}
\begin{split}
\mathcal{A}_1&=\text{Tr}\left[
(-\frac{\slashed{P}}{2}+\slashed{q}+m)\g^\a(\frac{\slashed{P}}{2}+\slashed{q}+m)\Pi(P,q,m)
\right]\,,
\end{split}
\end{align}
with
\begin{align}
\Pi(P,q,m)= (-i g T^{a_3}\g^{\mu_3})\frac{i(\slashed{k}_3 +
\frac{\slashed{P}}{2}+\slashed{q}+m)}{(k_3+\frac{P}{2}+q)^2-m^2}(-i
g T^{a_1}\g^{\mu_1})\frac{i(-\slashed{k}_2
-\frac{\slashed{P}}{2}+\slashed{q}+m)}{(-k_2-\frac{P}{2}+q)^2-m^2}
(-i g T^{a_2}\g^{\mu_2})\,.
\end{align}
As we study the large $p_T$ limit, both $P$ and $q$ have been
boosted to a similar direction, say $\hat P$, thus, they have the
following decomposition:
\begin{align}\label{eq:decom}
\begin{cases}
&P^\mu =\hat P^\mu + \lambda P_\bot^\mu + \lambda^2 \frac{P^2-P_\bot^2}{2 P^+} n^\mu\,,\\
&q^\mu =\frac{\zeta}{2}\hat P^\mu + \lambda q_\bot^\mu + \lambda^2
\frac{q^2-q_\bot^2}{2 q^+} n^\mu\,,
\end{cases}
\end{align}
where $\zeta=\frac{2q^+}{P^+}=\frac{2q\cdot n}{P\cdot n}$, $n$ is a
light like momentum which satisfies $(P\cdot n)^2/(n^0)^2 \gg P^2$,
and $\lambda$ is used to denote the power counting of the
corresponding term, that is, the term proportional to $\lambda^i$
behaviors as $O(M^i)$. Using the on-shell relations
$\left(\pm\frac{P}{2}+q \right)^2=m^2$ and the fact that
\begin{align}
(\pm\frac{\slashed{P}}{2}+\slashed{q}+m) \slashed{n}
(\pm\frac{\slashed{P}}{2}+\slashed{q}+m) = (\pm 1+\zeta)P^+
(\pm\frac{\slashed{P}}{2}+\slashed{q}+m)\,,
\end{align}
we can rewrite $\mathcal{A}_1$ as
\begin{align}
\begin{split}
\mathcal{A}_1&=\frac{-1}{(1-\zeta^2)P^{+2}}\text{Tr}\left[
(-\frac{\slashed{P}}{2}+\slashed{q}+m)\slashed{n}
\mathbf{1}(-\frac{\slashed{P}}{2}+\slashed{q}+m)
\g^\a(\frac{\slashed{P}}{2}+\slashed{q}+m) \mathbf{1} \slashed{n}
(\frac{\slashed{P}}{2}+\slashed{q}+m) \Pi(P,q,m) \right]\,,
\end{split}
\end{align}
where we also inserted two unit matrixes. Doing the Fierz
transformation
\begin{align}
\mathbf{1}_{ij}\mathbf{1}_{kl}=\frac{1}{4}\sum_\lambda
\Gamma^\lambda_{il}\Gamma_{\lambda,kj}\,,
\end{align}
with $\Gamma^\lambda=\mathbf{1}, \g^5, \g^\mu,\g^5\g^\mu,
\s^{\mu\nu}/\sqrt{2}$ and $\Gamma_\lambda=\Gamma^{\lambda\dag}$,
$\mathcal{A}_1$ becomes
\begin{align}
\begin{split}
\mathcal{A}_1&=\frac{-1}{4(1-\zeta^2)P^{+2}}\sum_\lambda\text{Tr}\left[
\Gamma_\lambda(-\frac{\slashed{P}}{2}+\slashed{q}+m)
\g^\a(\frac{\slashed{P}}{2}+\slashed{q}+m) \right]\\
&\qquad\qquad\qquad\qquad\times\text{Tr}\left[
(-\frac{\slashed{P}}{2}+\slashed{q}+m)\slashed{n}
\Gamma^\lambda\slashed{n} (\frac{\slashed{P}}{2}+\slashed{q}+m)
\Pi(P,q,m) \right]\\
&=\frac{-1}{2(1-\zeta^2)P^{+2}}\sum_{\lambda=1,2,3}\text{Tr}\left[
\hat\Gamma^\lambda(-\frac{\slashed{P}}{2}+\slashed{q}+m)
\g^\a(\frac{\slashed{P}}{2}+\slashed{q}+m) \right]\\
&\qquad\qquad\qquad\qquad\times\text{Tr}\left[
(-\frac{\slashed{P}}{2}+\slashed{q}+m)\hat
\Gamma^\lambda(\frac{\slashed{P}}{2}+\slashed{q}+m) \Pi(P,q,m)
\right]\,,
\end{split}
\end{align}
where relations
\begin{align}
\Gamma_\lambda \otimes \slashed{n} \Gamma^\lambda\slashed{n} =
0\,,~0\,,~ 2 \slashed{n} \otimes \slashed{n}\,, ~2 \g^5 \slashed{n}
\otimes \g^5 \slashed{n}\,, ~~2 \g^\mu_\perp \slashed{n} \otimes
\g^\mu_\perp \slashed{n}\,\nonumber
\end{align}
have been used, and
$\hat\Gamma^1=\slashed{n}\,,\hat\Gamma^2=\g^5\slashed{n}\,,
\hat\Gamma^3=\g^\mu_\perp \slashed{n}$. Note that we can think of
$\text{Tr}\left[
\hat\Gamma^\lambda(-\frac{\slashed{P}}{2}+\slashed{q}+m)
\g^\a(\frac{\slashed{P}}{2}+\slashed{q}+m) \right]$ to be $O(M)$
effectively, because squaring it will give a $O(M^2)$ result, thus,
we do leading power approximation for other terms, which gives
\begin{align}
\begin{split}
\mathcal{A}_1&=\frac{1}{8P^{+2}}\sum_{\lambda=1,2,3}\text{Tr}\left[
\hat\Gamma^\lambda(-\frac{\slashed{P}}{2}+\slashed{q}+m)
\g^\a(\frac{\slashed{P}}{2}+\slashed{q}+m) \right]\text{Tr}\left[
\hat{\slashed{P}}\hat \Gamma^\lambda \hat{\slashed{P}} \Pi(\hat
P,\frac{\zeta}{2} \hat P,0) \right] + O (M^2)\,,
\end{split}
\end{align}
with
\begin{align}
\Pi(\hat P,\frac{\zeta}{2} \hat P,0)= (-i g
T^{a_3}\g^{\mu_3})\frac{i(\slashed{k}_3 + \frac{1+\zeta}{2}
\hat{\slashed{P}})}{(k_3+\frac{1+\zeta}{2} \hat{P})^2}(-i g
T^{a_1}\g^{\mu_1})\frac{i(-\slashed{k}_2 - \frac{1-\zeta}{2}
\hat{\slashed{P}})}{(-k_2- \frac{1-\zeta}{2} \hat{P})^2} (-i g
T^{a_2}\g^{\mu_2})\,.
\end{align}
Similar to Sec.~\ref{sec:jpsiLO}, we choose an axial gauge so that
$\hat P \cdot \e(k_i)=0$, then some terms in $\Pi(\hat
P,\frac{\zeta}{2} \hat P,0)$ do not contribute because there is a
$\hat{\slashed{P}}$ on its either side. Thus,
\begin{align}
\begin{split}
\Pi(\hat P,\frac{\zeta}{2} \hat P,0)\sim & (-i g
T^{a_3}\g^{\mu_3})\frac{i \slashed{k}_3 }{(k_3+\frac{1+\zeta}{2}
\hat{P})^2}(-i g T^{a_1}\g^{\mu_1})\frac{-i \slashed{k}_2}{(-k_2-
\frac{1-\zeta}{2} \hat{P})^2} (-i g T^{a_2}\g^{\mu_2})\\
=& -i g^3 \frac{4}{1-\zeta^2} T^{a_3} T^{a_1} T^{a_2}
\g^{\mu_3}\frac{\slashed{k}_3 }{(k_3+\hat{P})^2}\g^{\mu_1}\frac{-
\slashed{k}_2}{(k_2+\hat{P})^2} \g^{\mu_2}\,,
\end{split}
\end{align}
and
\begin{align}
\begin{split}
\mathcal{A}_1&=\frac{ -i g^3}{2(1-\zeta^2)P^{+2}} \text{Tr}\left[
T^{a_1} T^{a_2} T^{a_3} \right] \sum_{\lambda=1,2,3}\text{Tr}\left[
\hat\Gamma^\lambda(-\frac{\slashed{P}}{2}+\slashed{q}+m)
\g^\a(\frac{\slashed{P}}{2}+\slashed{q}+m) \right]\\
&\qquad\qquad\qquad\times\text{Tr}\left[ \hat{\slashed{P}}\hat
\Gamma^\lambda \hat{\slashed{P}} \g^{\mu_3}\frac{\slashed{k}_3
}{(k_3+\hat{P})^2}\g^{\mu_1}\frac{- \slashed{k}_2}{(k_2+\hat{P})^2}
\g^{\mu_2} \right] + O (M^2)\,,
\end{split}
\end{align}
Observing that $\lambda=3$ does not contribute because there are an
odd number of Dirac matrices in the last trace, $\lambda=2$ does not
contribute because $\text{Tr}\left[
\hat\Gamma^2(-\frac{\slashed{P}}{2}+\slashed{q}+m)
\g^\a(\frac{\slashed{P}}{2}+\slashed{q}+m) \right] \propto
\e^{nPq\a}$, thus, $\mathcal{A}_1$ is odd in in $q$. As a result, we
get
\begin{align}
\begin{split}
\mathcal{A}_1&=\frac{ -i g^3}{2(1-\zeta^2)P^{+2}} \text{Tr}\left[
T^{a_1} T^{a_2} T^{a_3} \right] \text{Tr}\left[
\slashed{n}(-\frac{\slashed{P}}{2}+\slashed{q}+m)
\g^\a(\frac{\slashed{P}}{2}+\slashed{q}+m) \right]\\
&\qquad\qquad\qquad\times\text{Tr}\left[
\hat{\slashed{P}}\slashed{n} \hat{\slashed{P}}
\g^{\mu_3}\frac{\slashed{k}_3 }{(k_3+\hat{P})^2}\g^{\mu_1}\frac{-
\slashed{k}_2}{(k_2+\hat{P})^2}
\g^{\mu_2} \right] + O (M^2)\\
&=\frac{ -i g^3}{2(1-\zeta^2)P^{+}} \text{Tr}\left[ T^{a_1} T^{a_2}
T^{a_3} \right] \text{Tr}\left[
\slashed{n}(-\frac{\slashed{P}}{2}+\slashed{q}+m)
\g^\a(\frac{\slashed{P}}{2}+\slashed{q}+m) \right]\\
&\qquad\qquad\qquad\times\left\{ \text{Tr}\left[
\g^{\mu_3}\g^{\mu_1}\frac{\slashed{k}_2}{u} \g^{\mu_2} \right]
+\text{Tr}\left[ \g^{\mu_3}\frac{\slashed{k}_3 }{s}\g^{\mu_1}
\g^{\mu_2} \right] \right\} + O (M^2)\,.
\end{split}
\end{align}
Comparing it with Eq.~\eqref{eq:A1bar2} and using a similar
argument, we find $\mathcal{A}_1+ \mathcal{A}_2 + \mathcal{A}_3 = O
(M^2)$ and that $\frac{d\s}{dp_T^2}\propto M^4/p_T^8$ holds.

Combining the above two points, we know that, in the large $p_T$
limit, the denominator is $S_3^{-2}$ and the numerator has a factor
$S_1^2$. The rest numerator, as it is symmetric in $s$, $t$, and $u$
and has a mass dimension of $[M]^8$, must be proportional to $S_2^2$
using the general form in Sec.~\ref{sec:form}. Finally, we find that
the relativistic correction term is proportional to
$\frac{S_1^2S_2^2}{S_3^2}$, which is the same as the term at LO in
$v^2$. This behavior for next-to-leading order (NLO) relativistic
correction has been found in Ref.~\cite{Fan:2009zq}, but the
generalization to all orders in this work is new.

We conclude this section by explaining the logic to prove
$\frac{d\s}{dp_T^2}\propto M^4/p_T^8$. After the Fierz
transformation, we in fact factorize the amplitude to soft parts
[with scales of $O(M)$] and hard parts [with scales of $O(p_T)$].
This factorized form is equivalent to the double parton
fragmentation formula in Ref.~\cite{Kang:2011mg}, which gives the
contribution of $O(M^2)$ at the cross section level. Factorized
terms are then shown to vanish at the large $p_T$ limit. Therefore,
remaining terms can only give contribution at $O(M^4)$.

\section{Summary and outlook}\label{sec:summary}

In heavy quarkonium physics, many processes involve three gluons,
which result in $s$, $t$, $u$ symmetric cross sections or decay
widths. In this work we study the general form of $s$, $t$, $u$
symmetric polynomials, and find that they can be expressed as
polynomials of $S_1$, $S_2$, and $S_3$ where the symmetry is
manifest. For the massless case, the general form is summarized in
Eq.~\eqref{eq:massless}, and for the massive case, the general form
is summarized in Eq.~\eqref{eq:massive}. These general forms can be
used to simplify expressions that are symmetric in $s$, $t$, and
$u$. Asymptotically, this method can reduce the length of the
original expression to one-sixth. Based on these general forms, one
can also predict many interesting results by simple analysis. We
give two examples regarding $\jpsi$ hadron production in this work.
In the first example we work within the color-singlet model at LO in
$v^2$. By only arguing that the differential cross section has the
behavior $\frac{d\s}{dp_T^2}\propto M^4/p_T^8$ in large $p_T$ limit,
namely, $s,~t,~u\gg M^2$, we successfully reproduce the exact
differential cross section up to four unknown constant numbers. In
the second example, we consider relativistic corrections for
color-singlet model in large $p_T$ limit. By showing that, in large
$p_T$ limit, relativistic corrections do not change the denominator
and the behavior $\frac{d\s}{dp_T^2}\propto M^4/p_T^8$, we prove
that the differential cross section proportional to
$\frac{S_1^2S_2^2}{S_3^2}$ holds to all orders in $v^2$. This proof
not only explains the proportion relation at NLO in $v^2$ found in
Ref.~\cite{Fan:2009zq}, but also generalizes it to all orders.

Calculations of $O(v^2)$ and $O(\alpha_s)$ corrections to
heavy-quarkonium production and decay observables usually yield very
lengthy expressions. In view that the $s$, $t$, $u$ symmetry can
give so many constraints, it is possible to use our systematic
method to simplify these lengthy expressions and to exhibit their
symmetry. It will be also interesting to study symmetry induced by
four or even more gluons.

Proportion relations at NLO in $v^2$ are also found for the
color-octet channel~\cite{Xu:2012am}. However, since differential
cross sections for the color-octet channel have the behavior
$\frac{d\s}{dp_T^2}\propto M^2/p_T^6$ or $\frac{d\s}{dp_T^2}\propto
1/p_T^4$, proportion relations for them cannot be constrained by
only $s$, $t$, $u$ symmetry. We will study this problem in a
forthcoming work~\cite{maqiu}.

\begin{acknowledgments}
We would like to thank J.W. Qiu and Y.J. Zhang for useful
discussions. This work was supported by the U.S. Department of
Energy, under Contract No. DE-AC02-98CH10886. The Feynman diagrams
were drawn using Jaxodraw~\cite{Binosi:2008ig}.
\end{acknowledgments}

\input{main.arXiv}

\end{document}

%% file: main.arXiv
\providecommand{\href}[2]{#2}\begingroup\raggedright\endgroup